\documentclass [12pt]{article} 
\usepackage{amssymb,amsmath,amsthm}
\usepackage{graphicx}
\usepackage[utf8]{inputenc}
\usepackage[T1]{fontenc}

\newtheorem{thm}{Theorem}[section]
\newtheorem{prop}[thm]{Proposition}

\theoremstyle{remark}
\newtheorem{rem}[thm]{Remark}

\theoremstyle{definition}
\newtheorem{df}[thm]{Definition}
\newtheorem{ex}[thm]{Example}

\begin{document}

\title{{\bf New inconsistency indicators for incomplete pairwise comparisons matrices}} 

\author{Jacek Szybowski
\thanks{AGH University of Science and Technology, Faculty of Applied Mathematics, al. Mickiewicza 30, 30-059 Kraków, Poland, szybowsk@agh.edu.pl} \and
Konrad Ku\l{}akowski \thanks{AGH University of Science and Technology, The Department of Applied Computer Science, Al. Adama Mickiewicza 30, 30-059 Krakow, Poland, kkulak@agh.edu.pl} \and
Anna Prusak \thanks{Cracow University of Economics, The Department of Quality Management, ul. Rakowicka 27, 31-510 Krakow, Poland, anna.prusak@uek.krakow.pl}
}

\maketitle

\begin{abstract}
We introduce two new inconsistency measures for the incomplete pairwise comparisons matrices and show several examples of their calculation. We also carry out a comparative analysis of the new inconsistency indices with the existing ones based on the Monte Carlo simulation. 
\end{abstract}


\section{Introduction}
The pairwise comparisons method (also referred to as 'PC method')
is a process of comparing objects in pairs to judge which of them
is preferred over another. In PC technique only two elements at a
time are analysed. The first reported use of this method was electoral
system proposed in the 13th century by Ramon Llull, a medieval philosopher,
mathematician and theologian. This system was based on binary comparisons
of candidates \cite{Colomer2011rlfa}. The PC method was then improved
by a number of scholars such as e.g. Nicholas de Condorcet \cite{Condorcet1785esld,KASPER201936} and Louis L. Thurstone \cite{MaydeuOlivares2003otmf}. 

The PC methods appear very popular among decision makers. The reason
for that is simple. It is much easier to compare two elements than
a larger collection of alternatives at the same time. The most widely
known contemporary applications of the PC method are the Analytic
Hierarchy and Network Processes (AHP/ANP) and other multicriteria
decision support methods including ELECTRE, PROMETHE or MACBETH. The
AHP/ANP were proposed in 1960s by American mathematician, Thomas L.
Saaty. He first developed the AHP method, which is based on hierarchization
of a decision problem. A hierarchical model consists of the main decision
``goal'' (located on top of the hierarchy), ``criteria'', ``subcriteria''
and ``decision alternatives'' (bottom of the model) \cite{Saaty2000}.
The ANP method was designed as an extension of the hierarchy with
additional network-like connections. The analysis of both hierarchy
and network is based on the same mathematical principles which require
construction of square {\it pairwise comparisons matrices} (PCMs). Each PCM ($A$) reflects
judgments made within a group of homogenous elements. Individual values
of the matrix ($a_{ij}$) indicate the degree to which element $x_{i}$
is preferred over $x_{j}$ in relation to a parent criterion. For
each matrix priority vector is derived ($w=w_{1},w_{2}...,w_{n}$),
representing the ranking of elements according to their relative preference. 

Priorities (weights) derived for each PCM should be evaluated for
consistency. It reflects how precise and reliable decision makers
are in their subjective judgments. The term ,,consistency'' has
many definitions in the literature and is often associated with randomness
of pairwise comparisons \cite{Davvodi:2009wl} or rationality
of decision makers \cite{Gastes:2012dz,Brunelli:2015gy}. Consistency is seen as one of the main characteristics of data quality, along with accuracy, completeness and timeliness
\cite{Batini2009mfdq}. In Saaty's methods, consistency has mathematical dimension and is expressed by
the following condition: 

$a_{ik}=a_{ij}\cdot a_{jk}\forall i,j,k=1,\ldots,n$.

This condition means that each comparison in the matrix is confirmed
by any other comparison. In this way, inconsistency is understood
as a deviation from a perfectly coherent case and can be expressed
by specific coefficient. In \cite{Saaty1977asmf} Saaty developed a specific
measure for consistency which is called Consistency Index ($\textit{CI}$), and
its standardized version Consistency Ratio ($\textit{CR}$). A number of alternative
indicators of consistency can be found in the literature, e.g. in \cite{Aguaron2003tgci, Stein2007thci, Kulakowski2014tntb}. More detailed explanation of Saaty's indices and other consistency indicators is provided in Preliminaries section.

The role of consistency measures is to indicate whether a given PCM
is mathematically coherent, and therefore suitable for further analysis.
Thus, consistency indicator is seen as a criterion of acceptance or
rejection of the matrix. According to the Saaty's definition, PCM is consistent if $CR\leq0.1$ \cite{Saaty2000}. It has been often criticised for being too restrictive \cite{Apostolou1993}. Several algorithms of inconsistency reduction have been introduced. See, for example, \cite{Kulakowski2015acir,Koczkodaj2015fcod,KS2016}.

The existing consistency measures have been developed for complete PCMs only. However, in many cases we have to deal with partially filled PCMs, in which one or more comparisons are missing. A number of studies focused on methods determining the weights from incomplete matrices, but they do not propose relevant consistency indicators. Inconsistency of an incomplete PCM is treated as the inconsistency of its best, completely filled version \cite{Bozoki2010oocoSIMPL}. In this paper, we propose two new consistency indicators for incomplete PCMs. These indicators are based on the weight vectors induced by all the spanning trees of the graph related to a PCM.

\section{Preliminaries}
\subsection{Pairwise comparisons}

Given a finite set $A=\{a_{1},\ldots,a_{n}\}$ of alternatives we compare them pairwise, saving the results in a square $n \times n$ matrix $M$, called a PCM. The elements of such a matrix are positive with $1$s in the main diagonal. It is obvious that if an alternative $a_i$ is $x$ times better that $a_j$, then the latter is $x$ times worse than $a_i$. This leads us to a natural assumption of a PCM's {\it reciprocity}:
$$\forall i,j\ m_{ji}=\frac{1}{m_{ij}}.$$   

The main goal of the ranking computation procedure is to assign a positive weight $w_i$ to every alternative $a_{i}$. The ordered set of all the weights: 
$$
w=[w_{1},\ldots,w_{n}]^{T},
$$
is called {\it a weight (or priority) vector}.

One of the most popular methods of deriving the weight vector is the eigenvalue method (EVM) introduced in \cite{Saaty1977asmf}, which produces the weight vector as a normalized principal eigenvector.

Another way to obtain the ranking is the geometric mean method (GMM) introduced in \cite{Crawford1985anot}. By means of the logarithmic least square method it has been proved that the rescaled vector of geometric means of PCM rows may serve as the weight vector.

\begin{ex}
Consider a pairwise comparison matrix

\[
M=\left(\begin{array}{cccc}
1 & 2 & 3 & \frac{1}{6}\\
\frac{1}{2} & 1 & 5 & 1\\
\frac{1}{3} & \frac{1}{5} & 1 & \frac{1}{4}\\
6 & 1 & 4 & 1
\end{array}\right).
\]

Its principal eigenvalue equals $\lambda_{max}\approx4.677$ and
its principal eigenvector is given by
$$
w_{EV}=\left[0.43648,0.51571,0.13561,1\right]^{T}.
$$
The sum of its coordinates equals $2.0878$, so, after normalization,
we obtain a priority vector 
\begin{equation}\label{NEV}
w_{NEV}=\left[0.20906,0.24701,0.06495,0.47897\right]^{T}.
\end{equation}
This determines the order of alternatives: $a_{4},a_{2},a_{1},a_{3}$.

Similarly, using GMM, we compute the weight vector

$$
w_{GM}=\left[1,1.25743,0.3593,2.21336\right]^{T},
$$
which, normalized, takes the form
\begin{equation}\label{NGM}
w_{NGM}=\left[0.20704,0.26033,0.07439,0.45824\right]^{T}.
\end{equation}

As previously, the order of alternatives is: $a_{4},a_{2},a_{1},a_{3}$.
\label{EVGM}
\end{ex}

\subsection{Pairwise comparison graphs}

Fix a pairwise comparison matrix $M$.

\begin{df}{}
Let $G_M = (V;E;L)$ be a labelled, undirected graph with the set of vertices $V = \{a_1;\ldots; a_n\}$, the set of edges $E=\{\{a_i;a_j\}\subset V:\ i<j\}$, and the labelling function $L : E \longrightarrow \mathbb{R}$ so that $L(\{a_i; a_j\}) = m_{ij}$, for $i<j$.
The {\it graph} $G_M$ is said to be {\it induced by the matrix} $M$.
\end{df}

Let us recall that an undirected graph is {\it a (spanning) tree} if it is connected (i.e. there exists a path of edges connecting each two vertices) and includes no cycles (i.e. there's no path of pairwise different edges connecting a vertex with itself). Each spanning tree of an undirected graph with $n$ vertices contains exactly $n-1$ edges. 

\begin{rem}
It is a straightforward observation that a complete $n \times n$ PCM $M$ induces 
an udirected graph with $\frac{n(n-1)}{2}$ vertices. In the case of an incomplete matrix these numbers decrease. However, the lower limit of $G_M$'s edges which may allow to construct a priority vector is $n-1$. On the other hand, we must remember that this is a necessary but not sufficient condition.   
\end{rem}

As it was shown in \cite{KOCZKODAJ2015387} the necessary and sufficient condition to compare all alternatives is that a graph $G_M$ includes at least one tree. We will denote the set of all spanning trees of $G$ by $ST(G)$. According to \cite{Cayley}, the number of spanning trees of a graph with $n$ vertices equals $$NT(G)=n^{n-2}.$$

The Kirchoff’s Theorem \cite{Maurer1976} states that the number
of spanning trees in a connected graph $G$ with $n$ vertices
$a_1,\ldots,a_n$ coincidates with any cofactor of the Laplacian
matrix $L(G) = [l_{ij}]$ of $G$, whose elements are given by the
formula:
$$
l_{ij} = 
\left\{
\begin{array}{ll}
deg(a_i), & \mbox{if } i = j;\\
-1, & \mbox{if } i \neq j \mbox{ and } a_i \mbox{ is connected with } a_j;\\
0, & \mbox{otherwise}.
\end{array}\right.
$$

\begin{ex}
Consider a pairwise comparison matrix $M$ from Ex. \ref{EVGM} and its incomplete version obtained by removing $m_{13}$ and $m_{34}$.

\[
M'=\left(\begin{array}{cccc}
1 & 2 & ? & \frac{1}{6}\\
\frac{1}{2} & 1 & 5 & 1\\
? & \frac{1}{5} & 1 & ?\\
6 & 1 & ? & 1
\end{array}\right).
\]

The graphs $G_M$ and $G_{M'}$ induced by $M$ and $M'$ are shown on Fig. \ref{fig:graphs}.

\begin{figure}[t]
\begin{center}
\includegraphics[height=3.5in]{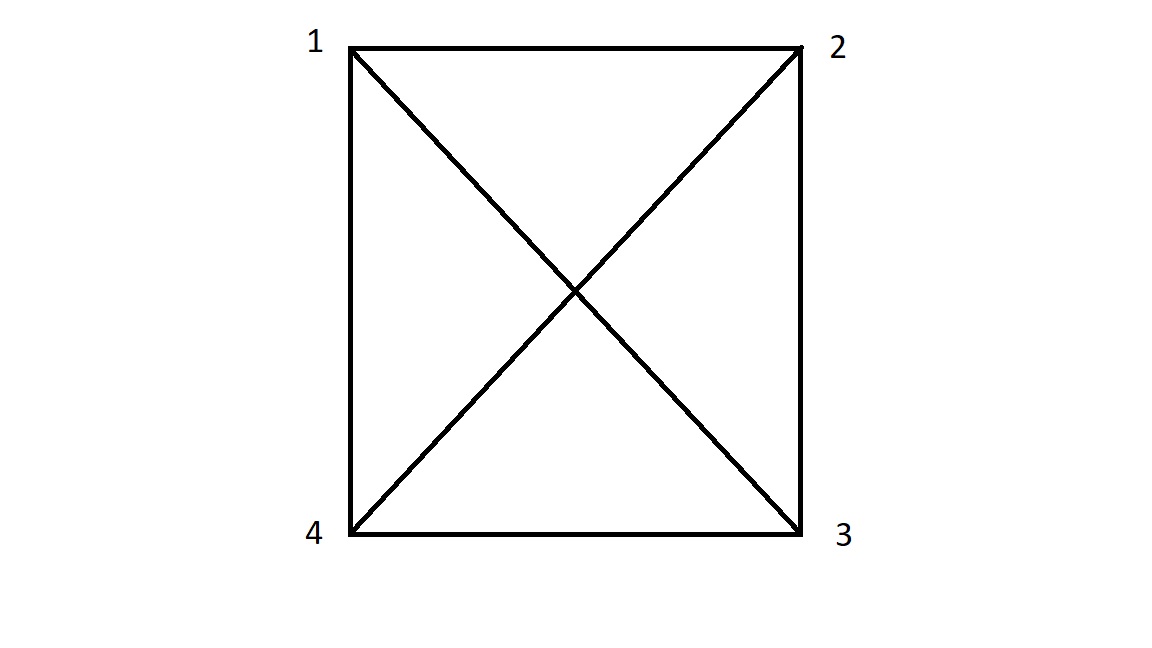}
\caption{The graphs of $M$ and $M'$.} \label{fig:graphs}
\end{center}
\end{figure}

Since $n=4$, the Cayley's Theorem implies that $NT(G_M)=16$.

The Laplacian matrix of $G_{M'}$ is as follows:

\[
L(G_{M'})=\left(\begin{array}{cccc}
2 & -1 & 0 & -1\\
-1 & 3 & -1 & -1\\
0 & -1 & 1 & 0\\
-1 & -1 & 0 & 2
\end{array}\right)
\]

Let us compute the cofactor of the left upper element of $L(G_{M'})$:

\[
L(G_{M'})_{11}=(-1)^{2}\cdot\left|\begin{array}{ccc}
3 & -1 & -1\\
-1 & 1 & 0\\
-1 & 0 & 2
\end{array}\right|=3.
\]
Thus, $NT(G_{M'})=3$. All spanning trees of $G_{M'}$ are illustrated on Fig. \ref{fig:trees}.

\begin{figure}[t]
\begin{center}
\includegraphics[height=3.2in]{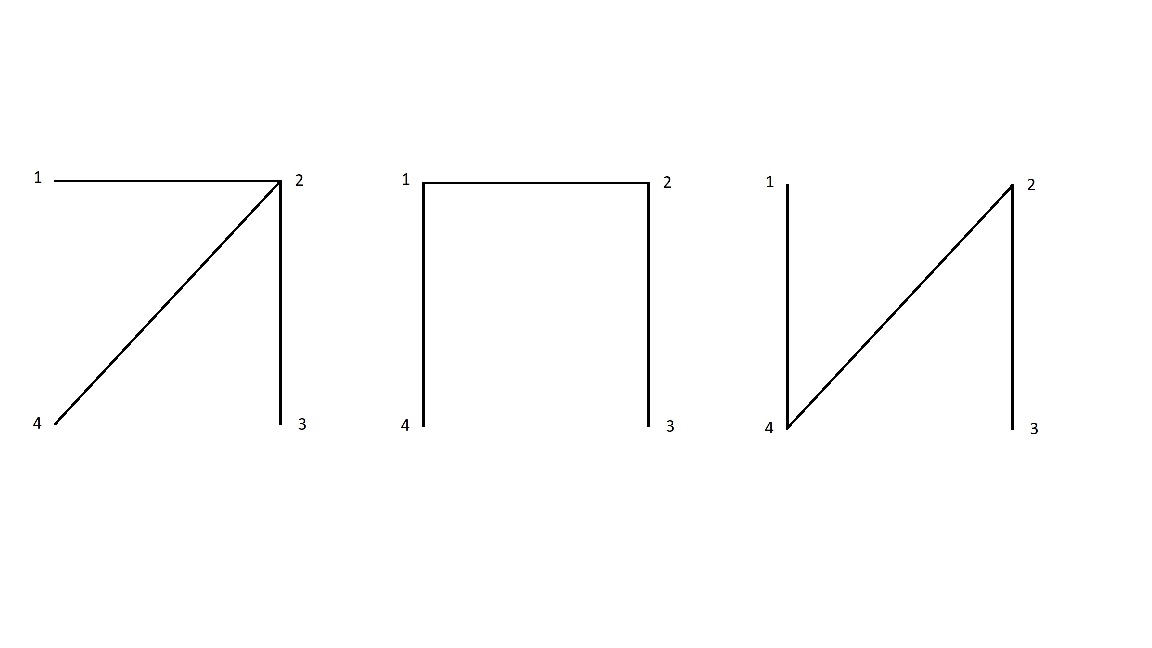}
\caption{The spanning trees of $G_{M'}$.} \label{fig:trees}
\end{center}
\end{figure}

\label{incompl}
\end{ex}

\section{Inconsistency}
A natural expectation concerning the PCMs is the transivity of pairwise comparisons. If, for example, alternative A is six times better than B, and B is twice worse than C, this should imply that A is three times better than C. Formally, we call a PCM $M$ {\it consistent} if 
$$\forall i,j,k \in \{1,\ldots,n\}\;\; a_{ij}a_{jk}a_{ki}=1.$$

In real applications consistent PCMs appear extremely rarely. Thus, in the literature there exist plenty of inconsistency measures. We recall some of them.
Let $M$ be a pairwise comparison $n \times n$ matrix.

\begin{df}{\cite{Saaty1977asmf}}
The Consistency Index of $M$ is given by
$$\textit{CI}(M)=\frac{\lambda_{max}-n}{n-1},$$
where $\lambda_{max}$ is the principal right eigenvalue of $M$ (i.e. the maximum one according to the absolute value).
\end{df} 

\begin{df}{\cite{Golden1989aamo}}
The $\textit{GW}$ Index of $M$ is defined as
$$\textit{GW}(M)=\frac{1}{n}\sum_{i=1}^n\sum_{j=1}^n\; |\bar{m}_{ij}-\bar{w}_i|,$$
where
$$\bar{m}_{ij}=\frac{m_{ij}}{\sum_{i=1}^n\; m_{ij}},$$
and
$$\bar{w}_i=\frac{w_i}{\sum_{k=1}^n\; w_k}$$
is a normalized weight vector obtained by EVM or GMM.
\end{df}

\begin{df}{\cite{Koczkodaj1993ando}}
The Koczkodaj inconsistency index of $M$ is given by the formula
$$K(M)=\max_{i<j<k}\min\left(\left|1-\frac{m_{ik}}{m_{ij}m_{jk}}\right|,\left|1-\frac{m_{ij}m_{jk}}{m_{ik}}\right|\right).$$
\end{df}

\begin{df}{\cite{Barzilai1998cmfp}}
The relative error of $M$ is equal to
$$\textit{RE}(M)=1-\frac{{\displaystyle \sum_{i=1}^n\sum_{j=1}^n}\; \left(\frac{1}{n} {\displaystyle\sum_{k=1}^n} \log m_{ik}-\frac{1}{n} {\displaystyle\sum_{k=1}^n} \log m_{jk}\right)^2}{{\displaystyle\sum_{i=1}^n\sum_{j=1}^n}\; \left(\log m_{ij}\right)^2}.$$
\end{df}

\begin{df}{\cite{Aguaron2003tgci}}
The Geometric Consistency Index of $M$ is defined as
$$\textit{GCI}(M)=\frac{2}{n-2}\sum_{i=1}^{n-1}\sum_{j=i+1}^n \ln^2\left(m_{ij}\frac{w_j}{w_i}\right),$$
where $w$ is a weight vector obtained by GMM.
\end{df}

\begin{df}{\cite{Stein2007thci}}
The Harmonic Consistency Index is given by
$$\textit{HCI}(M)=\frac{\left(\frac{1}{ \sum_{j=1}^n \frac{1}{ \sum_{i=1}^n m_{ij}}}-1\right)(n+1)}{n-1}.$$ 
\end{df}

\section{New measures of inconsistency}

\subsection{Manhattan index}
Let us consider two vectors $v$ and $w$ in $\mathbb{R}^n$. We define their {\it Averaged Manhattan Distance} as 
$$\textit{AMD}(v,w)=\frac{\sum_{i=1}^n |v_i-w_i|}{n}.$$
The above function may be naturally used as the measure of deviation of the vector weights obtained from the same PCM by two different methods.

\begin{ex}
The Averaged Manhattan Distance of the normalized weight vectors from Ex. \ref{EVGM} is equal $\textit{AMD}(w_{NEV},w_{NGM})=0.04551.$
\end{ex}

Consider a complete or incomplete PCM $M$ and its related graph $G_M$.
Every spanning tree $T$ of $G_M$ induces a unique normalized weight vector $w_T$. Denote the normalized geometric mean of all the vectors $w_T$ by $w_{GMT}(M)$. The derrivation of such a priority vector was proposed in \cite{Siraj2012east} and named as EAST (Enumerating All Spanning Trees). Let us recall that in the case of a complete PCM $w_{GMT}(M)$ coincidates with $w_{GM}(M)$ \cite{Lundy2016tmeo}, thus it is easy to calculate.

\begin{df}{}
A Manhattan Inconsistency Index ($\textit{MII}$) of a PCM $M$ is given by formula:
$$\textit{MII}(M)=\frac{\sum_{T \in ST(G_M)} \textit{AMD}(w_{GMT}(M),w_T)}{NT(G_M)}.$$
\end{df}

Obviously, a PCM matrix $M$ is consistent if and only if each spanning tree $T$ indicates the same normalized weight vector $w_T$, which coincidates with $w_{GMT}(M)$. This observation can be written as:

\begin{prop}{}
$$\textit{MII}(M)=0 \Leftrightarrow M \mbox{ is consistent}.$$ 
\end{prop}

\begin{ex}
Consider the PCM $M$ from Ex. \ref{EVGM} and the graph $G_M$. Fig. \ref{fig:span_trees} shows all its spanning trees (first row) and their corresponding weight vectors before and after normalization. The notation, for example, $4123$ corresponds to the tree where there is a path $a_4-a_1-a_2-a_3$, while $123+24$ denotes the tree with a path $a_1-a_2-a_3$ and an additional edge $a_2-a_4$.

\begin{figure}[t]
\begin{center}
\includegraphics[height=1.5in]{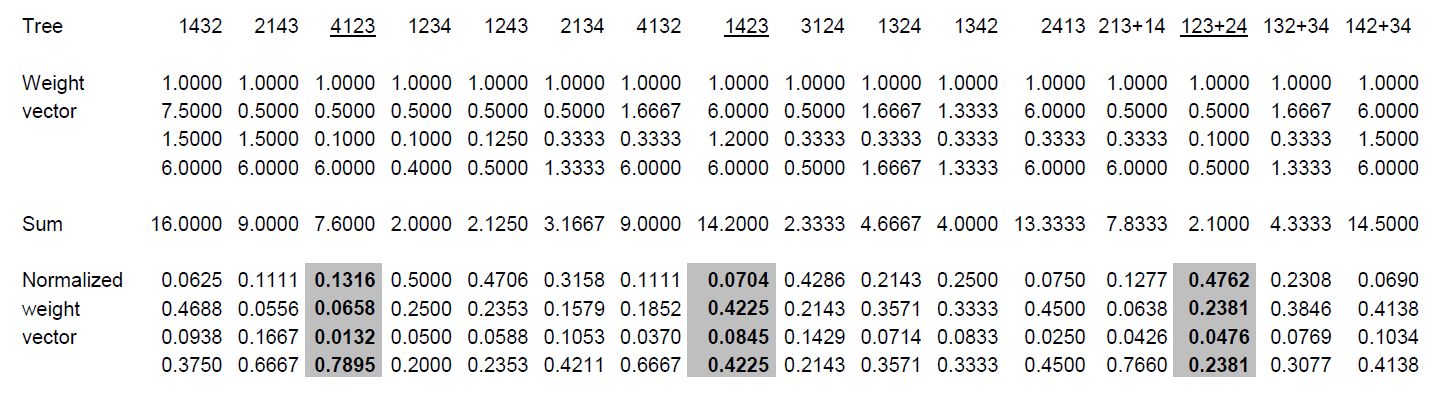}
\caption{The spanning trees and weight vectors of $G_{M}$.} \label{fig:span_trees}
\end{center}
\end{figure}

Since $NT(G_M)=16$ and $w_{GMT}(M)$ coincidates with (\ref{NGM}), we can calculate the Manhattan Inconsistency Index of $M$:
$$
\textit{MII}(M)=0.1111.
$$
\end{ex}

\begin{ex}
Now consider the PCM $M'$ from Ex. \ref{incompl} and its graph $G_M'$. The corresponding spanning trees are underlined in Fig. \ref{fig:span_trees}, while their normalized weight vectors are shaded. The resulting normalized weight vector is
\begin{equation}\label{NGT2}
w_{GMT}(M')=\left[0.2002,0.2292,0.0458,0.5247\right]^{T},
\end{equation}
so the Manhattan Inconsistency Index of $M'$ equals
$$
\textit{MII}(M')=0.1306,
$$
which differs only a little from $\textit{MII}(M)$.

\end{ex}
\subsection{Kendall index}
Obtaining a weight vector is a result of a process of decision making. However, in most cases a decision maker is satisfied with the information that one alternative is better than the other and they do not care by how much. Therefore, it is desirable to define an {\it order vector}, i.e. the vector assigning positions in a ranking to the alternatives. There is a simple rule how to obtain a ranking vector from a weight vector: the higher weight, the higher position in the ranking.

For example, the order vector corresponding to vectors given by (\ref{NEV}) and (\ref{NGM}) is
$$[3;2;4;1]^T.$$
In this case two methods produced the same vector. However, it often happens differently. Then we need a tool to compare by how much two rankings differ. A solution to this problem is a Kendall tau distance \cite{Kendall1938anmo,fagin}.

Let $p,q\in \{1,\ldots,n\}^{n}$ be two order vectors. We define their {\it Kendall tau distance} as
\begin{multline*}
K_{\textit{d}}(p,q)=
\#\left\{ (i,j)\,|\,\left(p_i>p_j\,\,\text{and}\,\,q_i<q_j\right)\right.\\
\left.\,\text{or}\,\left(p_i<p_j\,\,\text{and}\,\,q_i>q_j\right)\right.\\
\left.\,\text{or}\,\left(p_i=p_j\,\,\text{and}\,\,q_i\neq q_j\right)
\,\text{or}\,\left(p_i\neq p_j\,\,\text{and}\,\,q_i=q_j\right)\right\}.
\end{multline*}

\begin{ex}
Let $p=[3;2;4;1]^T$ and $q=[3;1;2;2]^T$ be two order vectors. Their Kendall tau distance equals $3$, since $p_1<p_3,$ while $q_1>q_3,$ $p_2>p_4,$ while $q_2<q_4,$ and $p_3>p_4,$ while $q_3=q_4.$
\end{ex}

\begin{rem}{}
$\forall p,q\in \{1,\ldots,n\}^{n}\;\;\; 0\leq K_{\textit{d}}(p,q) \leq \frac{n(n-1)}{2}$.
\end{rem}

Let $O:\mathbb{R}_{+}^{n}\rightarrow\{1,\ldots,n\}^{n}$
be the mapping assigning to every weight vector $w$ its order vector $O(w)$.

By analogy to the Manhattan Inconsistency Index we define the Kendall Inconsistency Index, which calculates the averaged Kendall tau distance of the order vectors induced by weight vectors of particular spanning trees and the order vector induced by their geometric mean.

\begin{df}{}
A Kendall Inconsistency Index ($\textit{KII}$) of a PCM $M$ is given by formula:
$$\textit{KII}(M)=\frac{\sum_{T \in ST(G_M)} K_{\textit{d}}(O(w_{GMT}(M)),O(w_T))}{NT(G_M)}.$$
\end{df}

\begin{ex}
Consider once more the PCM $M$ from Ex. \ref{EVGM} and the graph $G_M$. We have
$$O(w_{GMT}(M))=O(w_{GM}(M))=[3;2;4;1]^T.$$
The order vectors induced by weight vectors of all spanning trees of $G_M$ and their Kendall tau distance from $O(w_{GMT}(M))$ are illustrated in Fig. \ref{ovM}.

\begin{figure}[t]
\begin{center}
\includegraphics[height=1.1in]{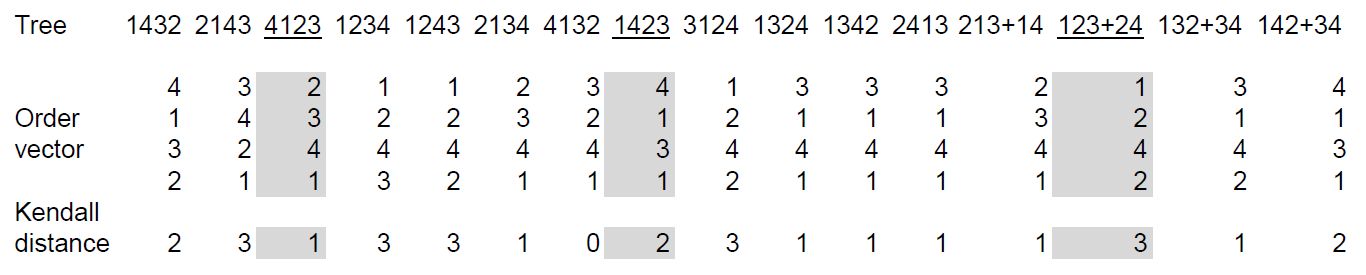}
\caption{The spanning trees, the order vectors and their Kendall tau distance from $O(w_{GMT}(M))$.} \label{ovM}
\end{center}
\end{figure}

Consequently,
$$\textit{KII}(M)=1.75,$$
which means that, on average, the orders of alternatives induced by different spanning trees differ from the orders induced by the whole PCM in less than two positions. 
\end{ex}

\begin{ex}
Now, let us consider again the PCM $M'$ from Ex. \ref{incompl} and its graph $G_M'$. The corresponding spanning trees are underlined in Fig. \ref{ovM}, while their order vectors and their Kendall tau distance from $$O(w_{GMT}(M'))=[3;2;4;1]^T$$ are shaded. 

As a result we get
$$\textit{KII}(M)=2.$$
\end{ex}

It is straightforward that
\begin{prop}{}
$$M \mbox{ is consistent} \Longrightarrow \textit{KII}(M)=0.$$ 
\end{prop}

However, the opposite implication is false.

\begin{ex}
Consider a pairwise comparison matrix

\[
M=\left(\begin{array}{cccc}
1 & 3 & 5 & 2\\
\frac{1}{3} & 1 & 2 & \frac{1}{2}\\
\frac{1}{5} & \frac{1}{2} & 1 & \frac{1}{3}\\
\frac{1}{2} & 2 & 3 & 1
\end{array}\right).
\]

Obviously, it is inconsistent, since, for example, $m_{12}m_{23}=6 \neq 5 = m_{13}$.
As we apply the GMM we get a normaized weight vector
$$
w_{NGM}=\left[0.48319,0.15688,0.08822,0.27172\right]^{T},
$$
and the corresponding order vector
$$O(w_{GMT}(M))=O(w_{GM}(M))=[1;3;4;2]^T.$$

Fig. \ref{trees0} shows all spanning trees of $G_M$, their corresponding weight vectors with their Manhattan distances from $w_{NGM}$, as well as the order vectors with the Kendall Tau distances from $O(w_{GMT}(M))$.

\begin{figure}[t]
\begin{center}
\includegraphics[height=2.2in]{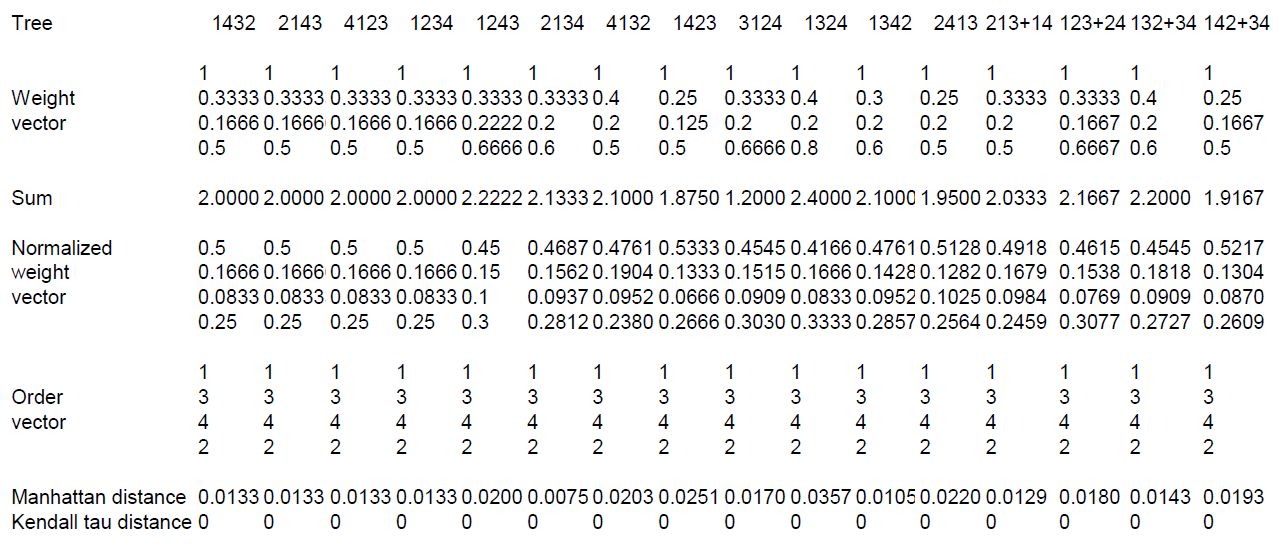}
\caption{The spanning trees and their corresponding weight and order vectors with their Manhattan and, respectively, Kendall Tau distances from the average ones.} \label{trees0}
\end{center}
\end{figure}

Let us notice that every spanning tree generates a weight vector slightly different than $w_{NGM}$. The resulting Manhattan Index equals 
$$
\textit{MII}(M)=0.0172,
$$
The other inconsistency indices of $M$ are also nonzero, although small:
\begin{center}

\begin{tabular}{|c|c|c|c|c|c|}
\hline
\textit{CI} & \textit{GCI} & \textit{HCI} & \textit{K} & \textit{GW} & \textit{RE}\\
\hline
0.005& 0.019& 0.004& 0.25& 0.064& 0.009\\
\hline
\end{tabular}

\end{center}

\vspace{.3cm}

However, all the order vectors induced by the spanning trees coincidate with $O(w_{GMT}(M))$, thus
$$\textit{KII}(M)=0.$$ \label{KII=0}
\end{ex}

Example \ref{KII=0} shows, that a zero value of Kendall Index does not imply full consistency. As the index may reach only a finite number of values, it splits the set of all pairwise comparison matrices into a finite number of classes.  This may be useful for classification of PCMs. 

In particular, we may define an {\it almost consistent matrix} as a PCM matrix, whose Kendall Inconsistency Index is equal to 0.

\section{The Monte Carlo analysis of the inconsistency indices correlation}

In order to compare different kinds of inconsistency indices we have prepared 30 series of thousand $5 \times 5$ PC matrices. The first series consists of 1000 fully consistent PC matrices derived from random vectors. The second series of matrices was created by multiplying each element above the main diagonals of random consistent PC matrices by a random number taken from the interval $[\frac{1}{2},2]$, which made them inconsistent. The successive series were created in a similar way but the multiplying factors belonging to intervals $[\frac{1}{3},3], [\frac{1}{4},4],\ldots,[\frac{1}{30},30]$, respectively. This resulted in more and more inconsistent (on average) matrices.

\begin{figure}[t]
\begin{center}
\includegraphics[height=3.5in]{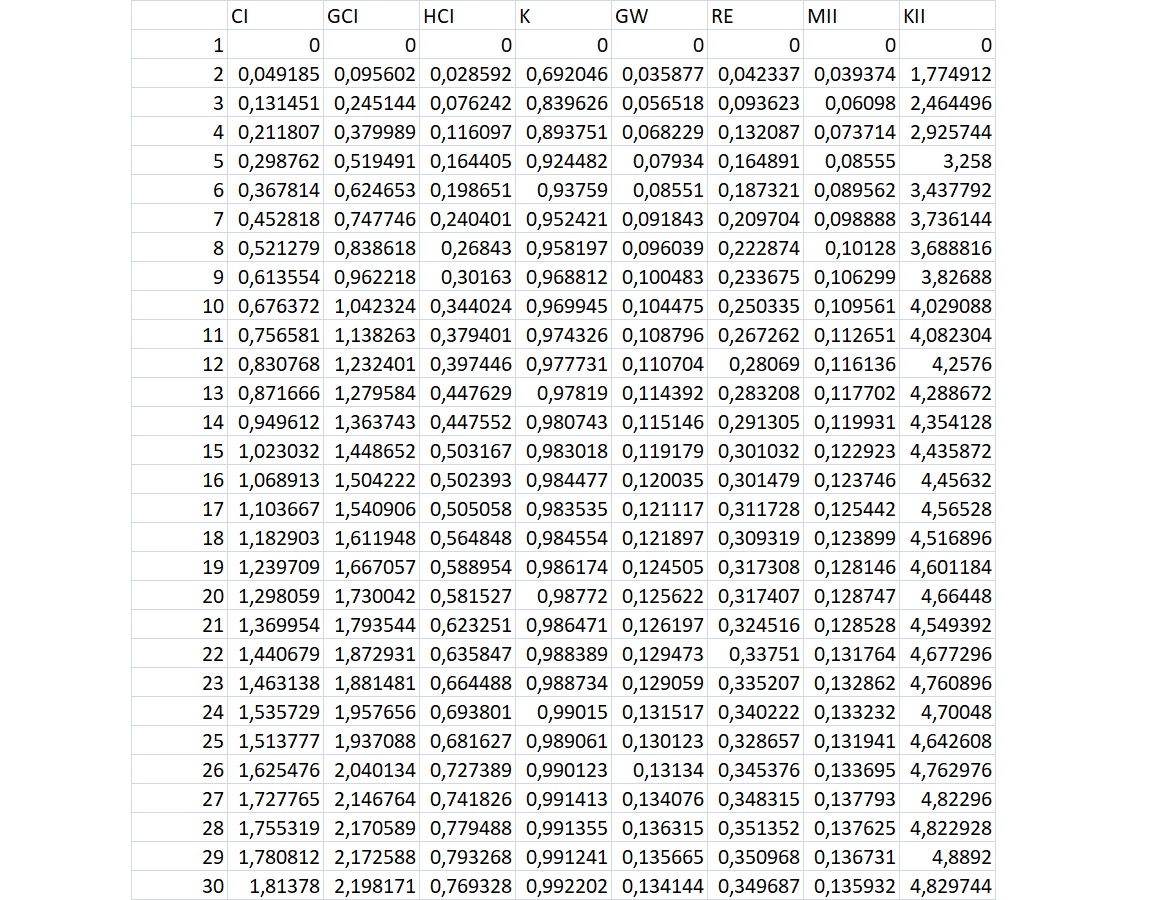}
\caption{The arithmetic means of inconsistency indices for random PC matrices.} \label{ind_num}
\end{center}
\end{figure}

\begin{figure}[t]
\begin{center}
\includegraphics[height=3.5in]{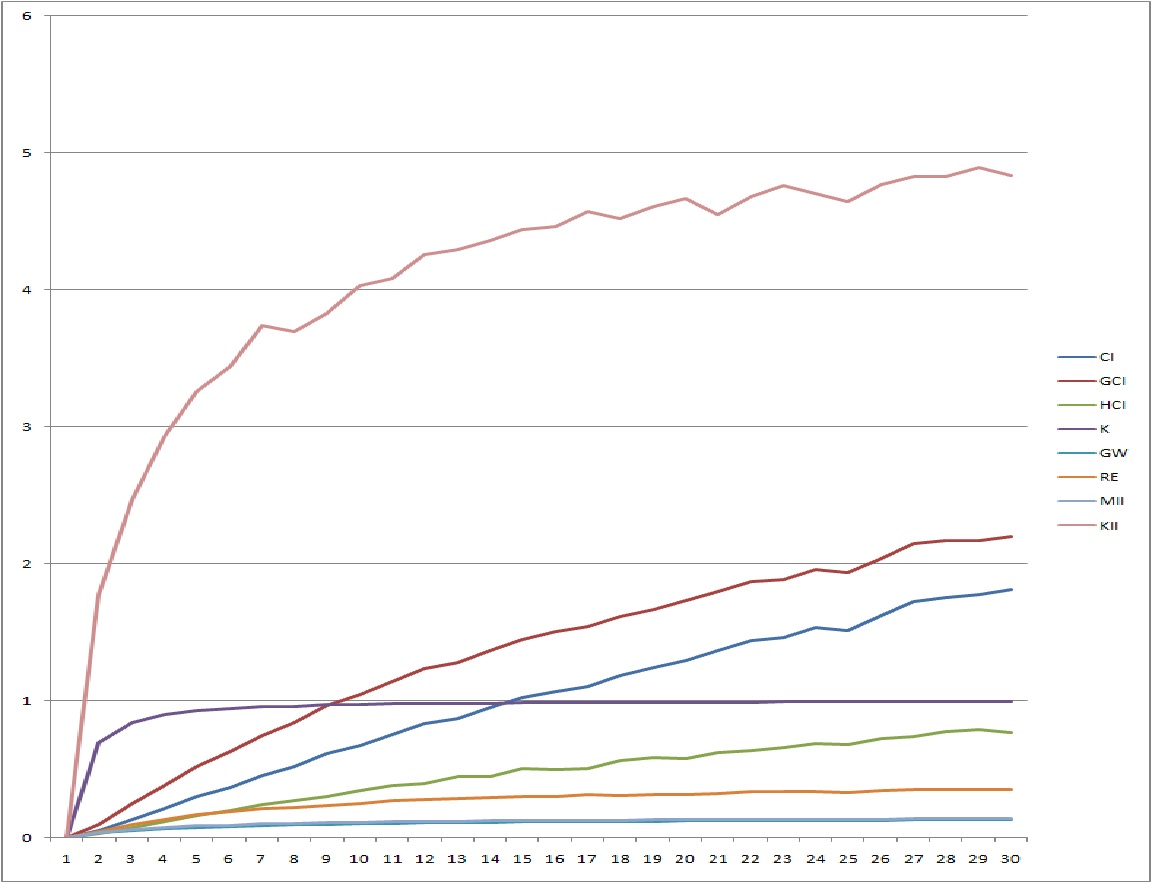}
\caption{The graphs of arithmetic means of inconsistency indices for random PC matrices.} \label{ind_graph}
\end{center}
\end{figure}


The next step was to calculate the $\textit{CI},\ \textit{GCI},\ \textit{HCI},\ \textit{K},\ \textit{GW},\ \textit{RE},\ \textit{MII}$ and $\textit{KII}$ indices for each PC matrix in each series. The arithmetic means of all the eight indices for each series has been presented in Fig. \ref{ind_num} and \ref{ind_graph}.

\begin{figure}[t]
\begin{center}
\includegraphics[height=3.0in]{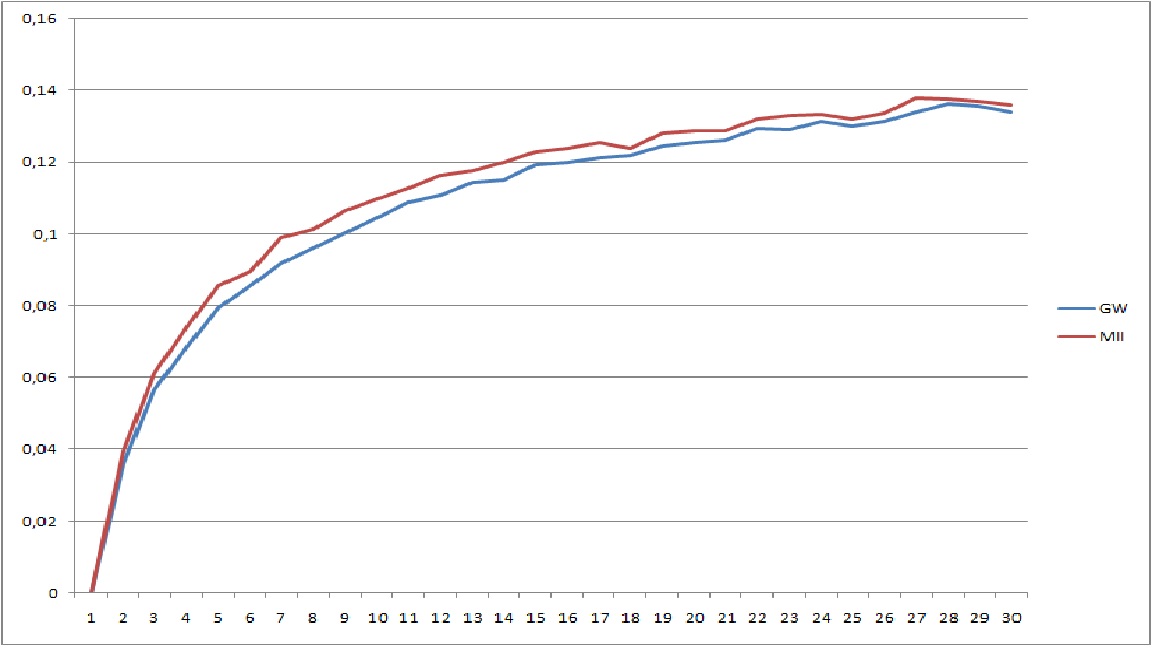}
\caption{The graphs of arithmetic means of $GW$ and $\textit{MII}$ indices for random PC matrices.} \label{GW+MII}
\end{center}
\end{figure}

The graphs of $GW$ and $\textit{MII}$ almost coincidate, which can be easily seen in Fig. \ref{GW+MII}.

\begin{figure}[t]
\begin{center}
\includegraphics[height=3.5in]{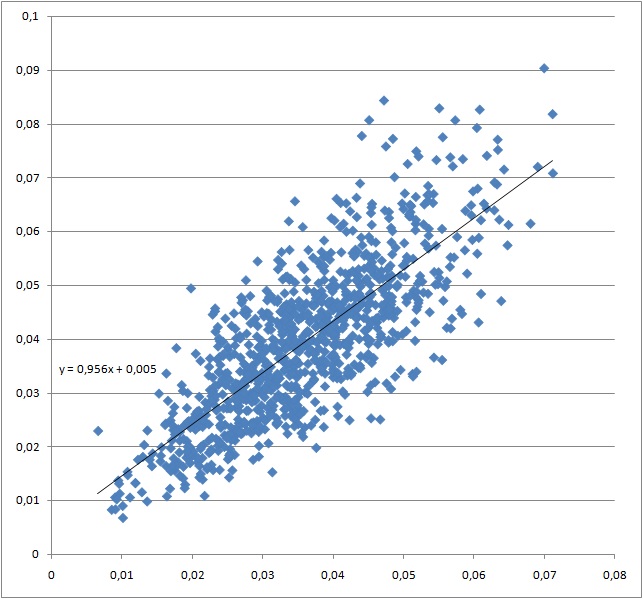}
\caption{The $GW$ and $\textit{MII}$ indices for random slightly inconsistent PC matrices.} \label{GW+MII_1}
\end{center}
\end{figure}

\begin{figure}[t]
\begin{center}
\includegraphics[height=3.5in]{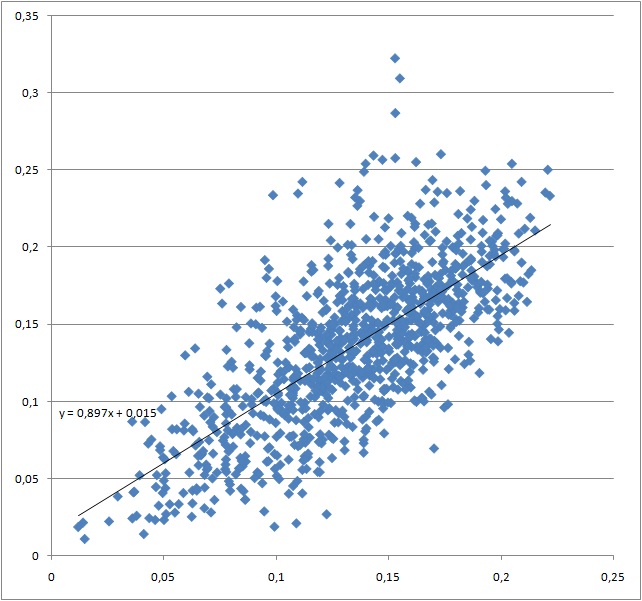}
\caption{The $GW$ and $\textit{MII}$ indices for random strongly inconsistent PC matrices.} \label{GW+MII_2}
\end{center}
\end{figure}

Fig. \ref{GW+MII_1} and \ref{GW+MII_2} show that for both for slightly (2nd series) and strongly (30th series) inconsistent random PC matrices their $GW$ and $\textit{MII}$ indices are almost equal. In the first case their correlation coefficient equals 0.782, while in the second case it is 0.709. Both are close to 1 which would mean the perfect coincidence.

\section{Conclusions}
We have proposed two new measures of inconsistency based on the spanning trees. Their advantage is the possibility to application in the case of incomplete PC matrices. As the Monte Carlo simulations have shown, the Manhattan Inconsistency Index and the Golden Wang Index give very similar results for complete pairwise comparisons matrices. we have also introduced the notion of almost inconsistent matrices, which may be used as the criterion of the input data acceptance.
  
\section{Acknowledgments}
The research is supported by The National Science Centre,
Poland, project no. 2017/25/B/HS4/01617 and by the Polish
Ministry of Science and Higher Education (task no. 11.11.420.004).


\bibliographystyle{plain}
\bibliography{papers_biblio_reviewed}

\end{document}